\pdfoutput=1
\documentclass[iop,twocolumn,jphysc,8pt,showpacs,floatfix,nofootinbib,
superscriptaddress
]{revtex4-1}
\usepackage{subfigure}
\usepackage{amssymb}
\usepackage{amsfonts}
\usepackage{amsmath}
\usepackage{amsthm}
\usepackage{epsfig}
\usepackage{graphicx}
\usepackage[usenames,dvipsnames]{color}
\usepackage[latin1]{inputenc}
\usepackage{hyperref}
\usepackage{comment}

\begin{document}

\title{Minimal domain size necessary to simulate the field enhancement factor numerically with specified precision}

\author{Thiago A. de Assis}
\email{thiagoaa@ufba.br}
\address{Instituto de F\'{\i}sica, Universidade Federal da Bahia,
   Campus Universit\'{a}rio da Federa\c c\~ao,
   Rua Bar\~{a}o de Jeremoabo s/n,
40170-115, Salvador, BA, Brazil}

\author{Fernando F. Dall'Agnol}
\address{Department of Exact Sciences and Education (CEE), Universidade Federal de Santa Catarina,
 Campus Blumenau, Rua João Pessoa, 2514, Velha, Blumenau 89036-004, SC, Brazil}
\email{fernando.dallagnol@ufsc.br}

\begin{abstract}

In the literature about field emission, finite elements and finite differences techniques are being increasingly employed to understand the local field enhancement factor (FEF) via numerical simulations. In theoretical analyses, it is usual to consider the emitter as isolated, i.e, a single tip field emitter infinitely far from any physical boundary, except the substrate. However, simulation domains must be finite and the simulation boundaries influences the electrostatic potential distribution. In either finite elements or finite differences techniques, there is a systematic error ($\epsilon$) in the FEF caused by the finite size of the simulation domain. It is attempting to oversize the domain to avoid any influence from the boundaries, however, the computation might become memory and time consuming, especially in full three dimensional analyses. In this work, we provide the minimum width and height of the simulation domain necessary to evaluate the FEF with $\epsilon$ at the desired tolerance. The minimum width ($A$) and height ($B$) are given relative to the height of the emitter ($h$), that is, $(A/h)_{min} \times (B/h)_{min}$ necessary to simulate isolated emitters on a substrate. We also provide the $(B/h)_{min}$ to simulate arrays and the $(A/h)_{min}$ to simulate an emitter between an anode-cathode planar capacitor. At last, we present the formulae to obtain the minimal domain size to simulate clusters of emitters with precision $\epsilon_{tol}$.  Our formulae account for ellipsoidal emitters and hemisphere on cylindrical posts. In the latter case, where an analytical solution is not known at present, our results are expected to produce an unprecedented numerical accuracy in the corresponding local FEF.

\end{abstract}

\maketitle

\section{Introduction}
\label{introduction}

Carbon nanotubes (CNTs) and carbon nanofibers (CNFs) have attracted great attention in cold field emission applications \cite{Heer1995, Minoux2005, Cole2015chapter, Cole}. The ability to design CNT or CNF field emitters with predictable electron emission characteristics allow their use as electron sources in various applications such as microwave amplifiers, electron microscopy, parallel beam electron lithography and advanced X-ray sources \cite{Cahay2014,Cole2015chapter}.

A common practice to model a CNT or CNF is take the emitter as a cylindrical classical conductor
capped with a conducting hemisphere, both of radius $r$. The total emitter's height
is $h$. This model is known as the ``hemisphere-on-cylindrical post" (HCP), have been extensively studied theoretically \cite{Edgcombe2001,Edgcombe,Xanthakis,RBowring,ZENG2009,Fuzinato,Unicamp2016} and is known to have an analytical counterpart of great complexity as compared with computational solution \cite{Edgcombe}.
In some studies, the hemi-ellipsoid model is preferred to represent the emitter for a couple of reasons: in this case, the the electrostatic potential distribution in the system has analytical solution and is a better representation of the emitter when the radius at the apex is much smaller than the radius at the base \cite{Edgcombe}.

The local field enhancement factor (FEF), particularly at the emitter's apex ($\gamma_a$), is an important quantity in field emission science. The apex-FEF facilitates the electron tunneling through the surface barrier and has a dramatic effect on the macroscopic emission current. As a consequence, the FEF makes the emission current very sensitive to the macroscopic electrostatic field, $E_M$ \cite{Miller,Miller2}. For a single-tip field emitter, the relevant characteristic barrier field at the apex $E_a$ may be formally related to $E_M$ by a characteristic apex-FEF \cite{Forbes2012b}, defined as follows:

\begin{equation}
 \gamma_a = \frac{E_{a}}{E_M}.
 \label{AFEF}
\end{equation}
Various physical causes of field enhancement can exist, but the more simple assumption is consider the existence of a conducting microprotrusion or nanoprotrusion at the emitter surface \cite{Edgcombe}. The question then arises of how the local apex-FEF can be accurately estimated.

Simulations that involve the solution of Laplace's equation, whether using finite elements or finite differences, require a minimum volume surrounding the region of interest. The minimum domain size (MDS) depends on the desired precision. If the boundaries are inadvertently close, they will affect the calculated electrostatic potential and the FEF may not respect the desired precision. Then, it might be attempting to overestimate the size of the domain. However, this procedure may become seriously time and memory consuming as the simulated systems become more demanding, mainly in three dimensional (3D) and/or in time-dependent models. Our analysis considers a two dimensional (2D) axisymmetric system, but it is also valid for 3D models, as we will discuss. Hereafter, we adopt the following notation regarding the dimensions of the domain: (i) the variables $A$ and $B$ are the width and height of the simulation domain as shown in Fig. \ref{Figbound}; (ii) the $A$ and $B$ relative to the height $h$ are much better parameters for our analysis. In fact, $A/h$ and $B/h$ are the variables we shall use to determine the systematic error; (iii) the $(A/h)_{min}$ and $(B/h)_{min}$ are the minimum dimensions that induces the tolerated systematic error $\epsilon_{tol}$ in the apex-FEF for single tip field emitters.

We start evaluating the minimum width and height $(A/h)_{min} \times (B/h)_{min}$ of the simulation domain, for single tip field emitter in an semi-infinite space. Then, we discuss why the dimensions found for a single tip field  emitter is also valid for emitters in an infinite array and for isolated emitters in a capacitor configuration. Finally, we provide a few numerical values to compare the predictions from our results and the actual simulations.
In most cases, the knowledge of $\left[(A/h)_{min},(B/h)_{min}\right]$ is convenient to choose the adequate size of the simulation domain that yields an error of $\sim1$\%, which can be considered good in view of the large uncertainties in FE experiments. However, for some analyses, the precision requirement is much larger and must be known. As an example, recently, Forbes has shown that the rate at which the FEF from a pair of spherical emitters tend toward the FEF of a single sphere is given by a power law decay as a function of the separation center to center \cite{RFJAP2016}. Prior to his work, many authors were assuming an exponential-type decay \cite{Groning2000,Groning1,Bonard,Jo,Harris2015AIP,Harris2016}. This power law was an important realization in the physical mechanism governing the FEF. Since then, the same power law behavior was observed in several different types of emitters like hemispheres on cylindrical posts and ellipsoidal emitters \cite{ForbesAssis2017,JPCM2018,arxiv2018,arxiv20182}, provided that the analysis has high precision, as we demonstrated in a previous work \cite{JPCM2018}.

This paper is organized as follows. In Section \ref{SP}, the simulation procedures, by using finite element method, are discussed. The results of the MDS, including hemi-ellipsoid single tip field emitter, arrays and capacitor configuration systems, hemisphere on a cylindrical post emitter and full three dimensional models of clusters are discussed on Sec.\ref{RD}. Section \ref{Conc} summarizes our conclusions.

\section{Simulation procedure}
\label{SP}

\begin{figure}[h!]
\includegraphics [width=8.5cm,height=7.2cm] {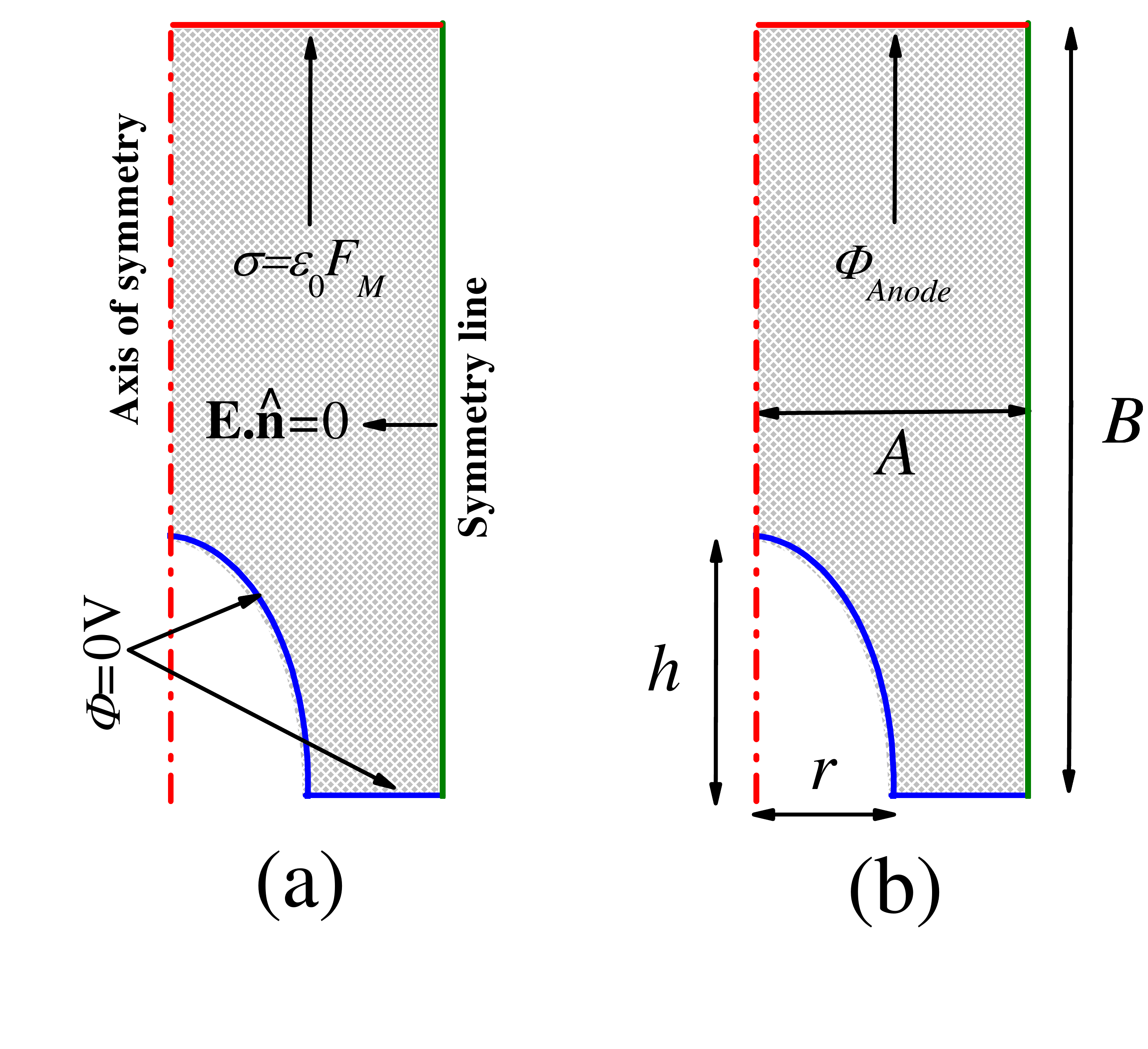}
\caption{Representation of the two most common physical systems to simulate field emission in the literature with the boundary conditions indicated. In (a) the top boundary is a Neumann boundary condition that imposes a vertically aligned electrostatic field as if the anode were much farther. In (b) the top boundary represents the anode with a Dirichlet boundary condition $\Phi=\Phi_{Anode}$.} \label{Figbound}
\end{figure}

Figure \ref{Figbound} represents the geometries of the physical systems we are interested in minimizing the size. It is a 2D axisymmetric system with a central emitter. We start analyzing an ellipsoidal emitter, for which $\gamma_{a}$ is known analytically \cite{Edgcombe}. Afterward, we shall discuss what errors to expect for emitters like hemispheres on posts and floating spheres. We assume the emitter to be perfectly conductor with no electric field penetration; hence, the interior of the emitter is removed from the simulation domain. The boundary conditions (BCs) on the emitter's surface and the bottom emitter's surface are grounded ($\Phi=0V$). The right hand side boundary is a symmetry line, i.e., this BC imposes the electrostatic field to be perpendicular to the normal vector from this boundary line ($\mathbf{E}.\mathbf{\hat{n}}=0$).

Two distinct BCs are common at the top boundary: Fig. \ref{Figbound}(a) represents a single tip field emitter. In this case, the top boundary is set as a surface charge density $\sigma = \varepsilon_0 E_M$, where $\varepsilon_0$ is the permittivity of vacuum. This BC assumes that the anode is much farther than the boundary itself. This is the BC we recommended when assuming that the anode is at infinity, as we shall discuss further. Figure \ref{Figbound}(b) represents a single tip field emitter with the top boundary representing the anode with a defined voltage $\Phi_{Anode}$.

We have used the commercial software COMSOL® v.5.3 based on the finite elements method to calculate the apex-FEF. The software evaluates the electrostatic potential distribution and the consequent electrostatic field in the domain. The apex-FEF calculated numerically, $\gamma_n$ is defined as the maximum value of the electrostatic field normalized by the applied field as shown in Eq. (\ref{AFEF}). Here the index ``$n$" indicates a numerical evaluation. We define the error in $\gamma_n$ relative to the analytical values of the field enhancement factor known for hemi-ellipsoidal emitters \cite{Edgcombe}:

\begin{equation}
 \epsilon = \frac{\left| \gamma_{a}- \gamma_{n}\right|}{\gamma_{a}} = \left| 1 - \frac{\gamma_n}{\gamma_a}\right|.
 \label{error}
\end{equation}
We want to stress that the analysis in this paper is not limited to any particular method (like finite elements) or computer code. Any computer code will require a MDS to yield a desired precision regardless the numerical precision used in the method. In our analyses, we took care to have enough numerical precision and sufficient number of elements not to compromise the evaluation of $\epsilon$, which is solely due to the finite size of the domain. In these analyses, the finite elements were concentrated where the electric field were higher and the number of elements were increased until the solution converged to the necessary precision for our purposes.
Although the Boundary Elements Method \cite{RBowring} is a better alternative to simulate the FEF, it does not apply in many cases. The MDS we provide here can be useful where the knowledge of the solution in the bulk of the system is necessary. Some examples are: space charge effect, mechanical oscillations in a field emission system or particle tracing analyses.

\section{Results and Discussion}
\label{RD}

\subsection{Hemi-Ellipsoid single tip field emitter}
\label{STFE}

\begin{figure}[h!]
\includegraphics [width=8.8cm,height=5.8cm] {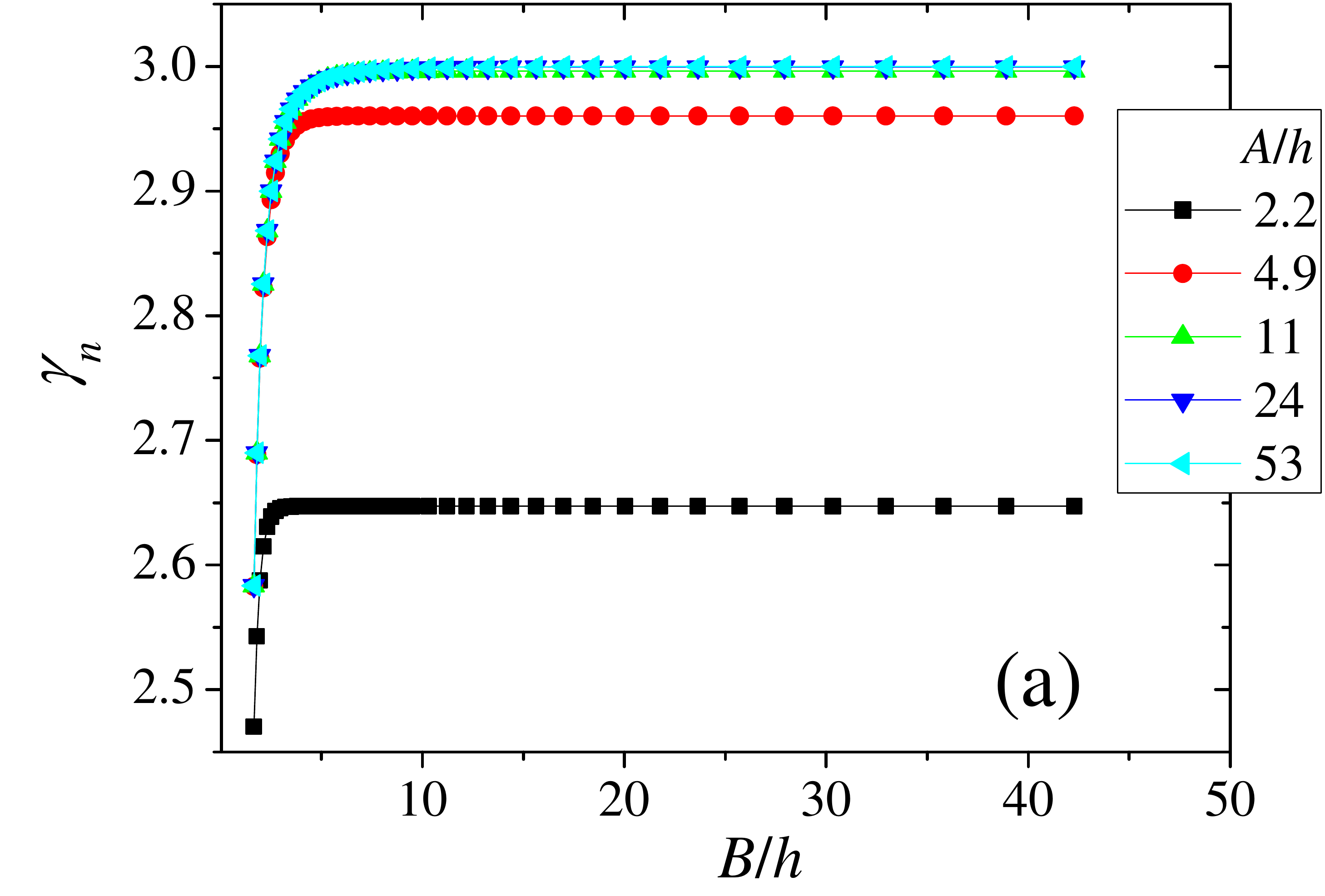}
\includegraphics [width=8.0cm,height=5.8cm] {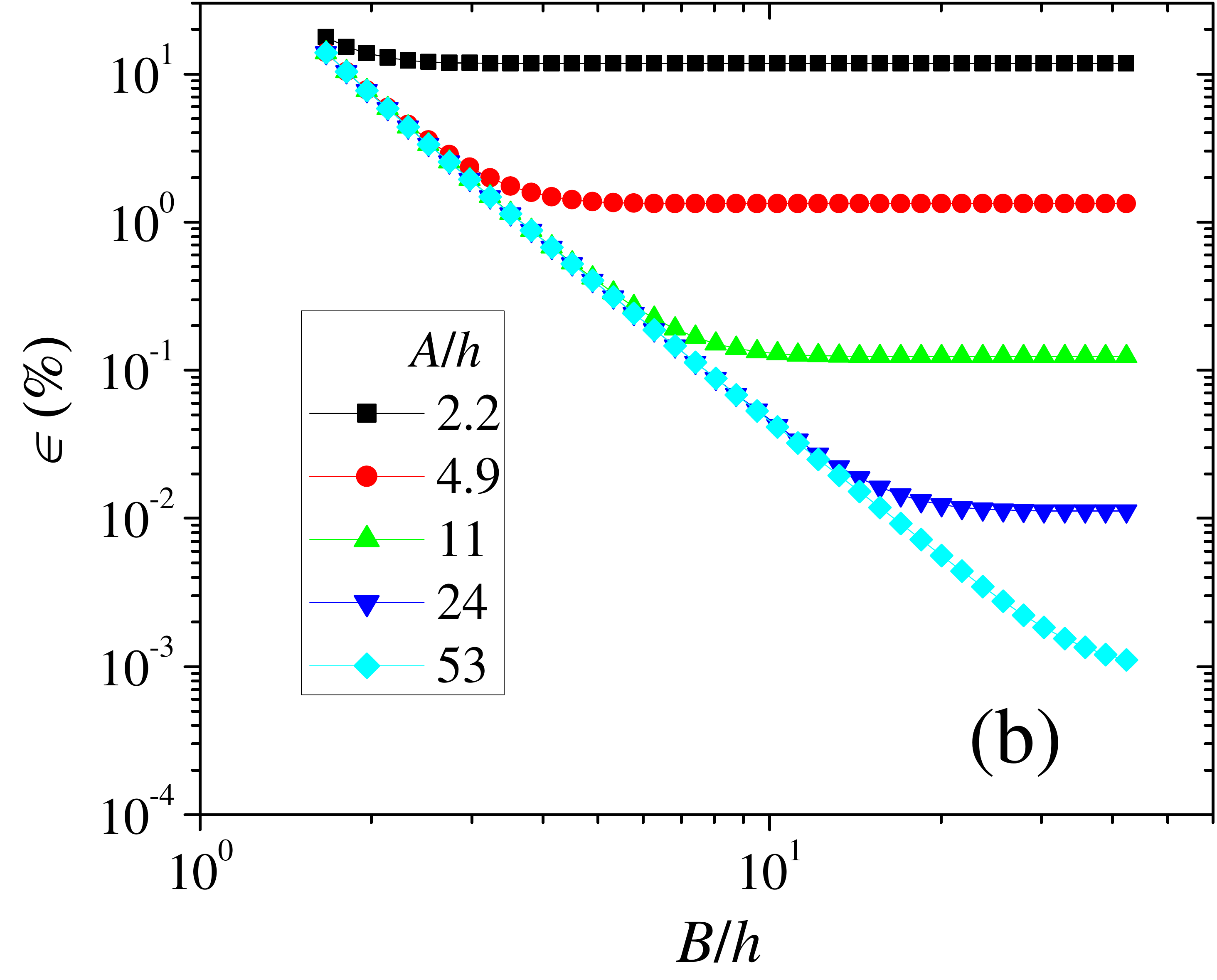}
\caption{Apex FEF and its systematic error due to the finite size of the simulation domain. In (a), the FEF converges its analytical value ($\gamma_n \rightarrow 3$) as the height and width of the simulation domain increases. In (b), as the height increases the error diminishes until the influence from the finite width saturates the error.} \label{Fighe}
\end{figure}

Figure \ref{Fighe}(a) illustrates a simple case for a hemispherical emitter, i.e. $h/r=1$, where $r$ is the radius of the emitter. The $\gamma_n$ converges to its analytical value ($\gamma_n \rightarrow 3$) as the width and height of the domain increases. Figure \ref{Fighe}(b) shows the corresponding systematic error obtained as the height is increased for several normalized widths $A/h$. In Fig. \ref{Fighe}(b) there is the systematic error $\epsilon$ as a function of the same variables shown in Fig. \ref{Fighe}(a). The $\epsilon$ decays linearly (in log-log scale) until the electrostatic influence from the top boundary becomes smaller than the influence from the lateral boundary and   starts to saturate.

\begin{figure}[h!]
\includegraphics [width=8.0cm,height=5.8cm] {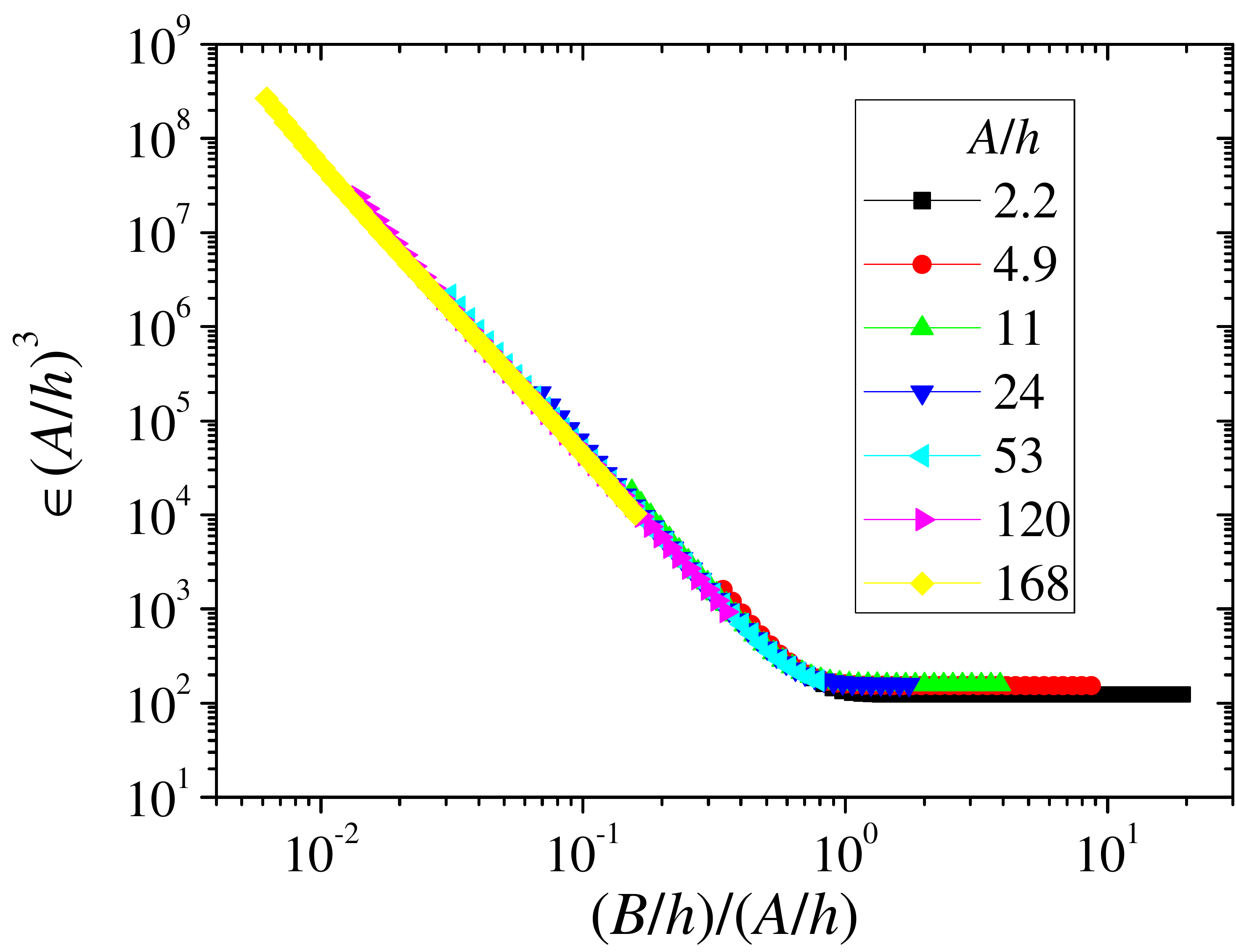}
\caption{Axes of Fig. \ref{Fighe}(b) rescaled to collapse all curves. The leftmost point in the plateau is the minimum domain size for all $\epsilon_{tol}$ and $A/h$. This point defines $(A/h)_{min}$ and $(B/h)_{min}$.} \label{Fighed}
\end{figure}
Figure \ref{Fighe}(b) can provide the MDS for a few values of  $\epsilon_{tol}$. As an example: Fig. \ref{Fighe}(b) shows that the curve for $A/h=11$ has  $\epsilon=0.12\%$ at the saturation, which extends for $B/h>10$. Hence, if ones tolerance happens to be $\epsilon_{tol}=0.12\%$, then the leftmost point in the plateau of the curve provide the MDS with $(A/h)_{min}=10$ and $(B/h)_{min}=11$. Better yet, instead of inspecting Fig.\ref{Fighe}(b) for the MDS, we managed to get a good collapse of all curves by plotting the variables $\epsilon (B/h)^3 \times (A/h)/(B/h)$ as shown in Fig.\ref{Fighed}. This is the most important result to obtain the $[(A/h)_{min},(B/h)_{min}]$ valid for all  $\epsilon_{tol}$. In Fig.\ref{Fighed}, a universal point of minimum can be obtained at the leftmost point in the plateau of the curves. The definition of this point depends on a criterion. In this work we assume $(A/h)_{min}$ to be 10\% higher than the value the end of the plateau, which gives $\epsilon_{tol} \left[(A/h)_{min}\right]^{3}\approx175$ and $\left[(B/h)_{min}/(A/h)_{min}\right]\approx0.83$. Then,

\begin{equation}
 (A/h)_{min} \approx \frac{5.59}{\sqrt[3]{\epsilon_{tol}}},
 \label{Ahmin}
\end{equation}
and
\begin{equation}
 (B/h)_{min} \approx \frac{4.64}{\sqrt[3]{\epsilon_{tol}}}.
 \label{Bhmin}
\end{equation}

\begin{figure}[h!]
\includegraphics [width=8.0cm,height=5.8cm] {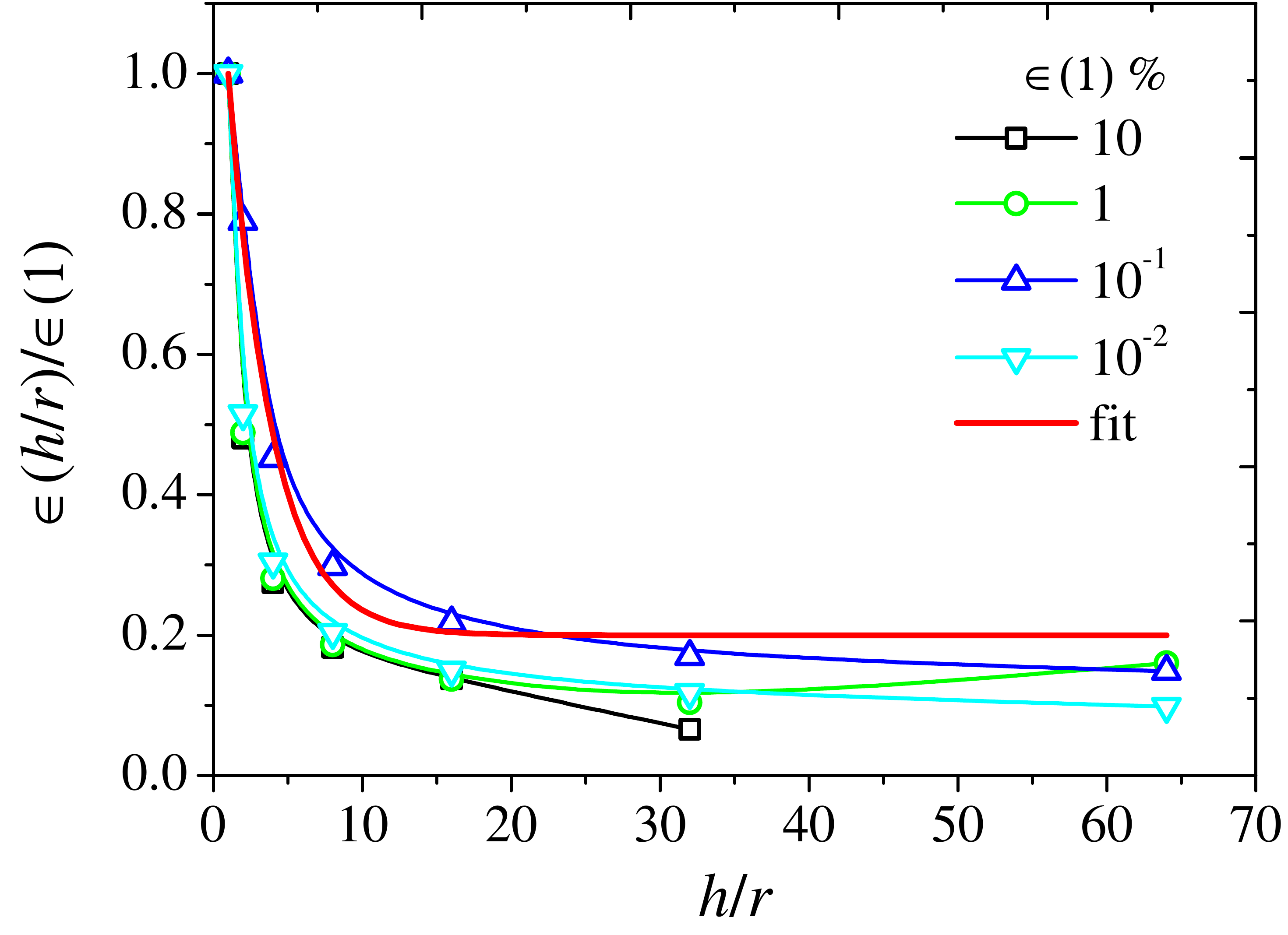}
\caption{Dependence of the systematic error with the aspect ratio. Given a domain size, the error for larger aspect ratios is sensitively smaller than the error for a unitary aspect ratio. The dependence of $\epsilon$ with $h/r$ can be taken into account to improve the minimum domain size.} \label{effhr}
\end{figure}

 Equations (\ref{Ahmin}) and (\ref{Bhmin}) are the upper bound of the MDS needed to simulate the apex-FEF with desired error $\epsilon_{tol}$ for a unitary aspect ratio. For higher aspect ratios, the MSD is considerably smaller. Figure \ref{effhr} shows how $\epsilon$  decreases as $h/r$ increases tending to saturate for $h/r\approx20$. Our analyses show that for any value of $A/h$ or $B/h$, the  $\epsilon(h/r)/\epsilon(1)$ can be modeled as

 \begin{equation}
 \frac{\epsilon(h/r)}{\epsilon(1)} \approx 0.2 + 0.8\exp\left[-0.345\left(h/r - 1 \right) \right].
 \label{ARM}
\end{equation}

Equation (\ref{ARM}) indicates that the error can drop 80\% as $h/r$ increases compared with a unitary aspect ratio. The $\epsilon_{tol}= \epsilon(h/r)$ is the error one is interested in having [not $\epsilon_{tol}= \epsilon(1)$]. Hence, we can replace $\epsilon(1)$ from Eq. (\ref{ARM}) in Eqs. (\ref{Ahmin}) and (\ref{Bhmin}) to incorporate the dependence with $h/r$ in the MDS as follows:

\begin{equation}
 (A/h)_{min} \approx 5.59 \times \sqrt[3]{\frac{0.2 + 0.8\exp\left[-0.345\left(h/r - 1 \right) \right]}{\epsilon_{tol}}},
 \label{Ahmin2}
\end{equation}
and
\begin{equation}
 (B/h)_{min} \approx 4.64 \times \sqrt[3]{\frac{0.2 + 0.8\exp\left[-0.345\left(h/r - 1 \right) \right]}{\epsilon_{tol}}}.
 \label{Bhmin2}
\end{equation}
Equations (\ref{Ahmin2}) and (\ref{Bhmin2}) predict the MDS much improved as compared to Eqs.(\ref{Ahmin}) and (\ref{Bhmin}).

\subsection{Arrays and capacitor configuration systems}

Our 2D system, as illustrated in Fig. \ref{Figbound}, can be used with good approximation to simulate a lattice as described in detail in Ref.\cite{Fuzinato}. The symmetry boundary acts as a mirror, so the emitter experiences a screening effect due to its own image, similar to the screening in a lattice, where the distance to the neighboring emitters is $L=2A$. Equations (\ref{Ahmin2}) and (\ref{Bhmin2}) apply straightforwardly to simulate arrays or an isolated emitter in a capacitor configuration. However, it is necessary to interpret $\epsilon_{tol}$  as the actual tolerated systematic error, not the error defined in Eq. (\ref{error}). For example, if one is interested in simulating a compact array with $\epsilon_{tol}=1\%$, then the electrostatic influence form the lateral boundary (which is fixed to $A=A_{array}$) causes $\epsilon$, as defined in Eq. (\ref{error}), to be larger than 1\%. However, the electrostatic influence from the symmetry boundary is not an error in the calculations; it is necessary to compute the fractional reduction in the FEF in arrays. The only electrostatic influence that has to be avoided is the influence from the top boundary, which is already granted to be less than 1\% if $(B/h)_{min}$ from Eq.(\ref{Bhmin2}) is respected. The same is valid for a capacitor configuration. In this case, the influence from the top boundary is necessary to compute the FEF correctly. However, for as long as $(A/h)_{min}$ from Eq. (\ref{Ahmin2}) is calculated with the desired tolerance $\epsilon_{tol}$, then the error from the lateral boundary will be smaller than $\epsilon_{tol}$.

\subsection{Hemisphere on a cylindrical post emitter - HCP model}

\begin{figure}[h!]
\includegraphics [width=8.0cm,height=5.8cm] {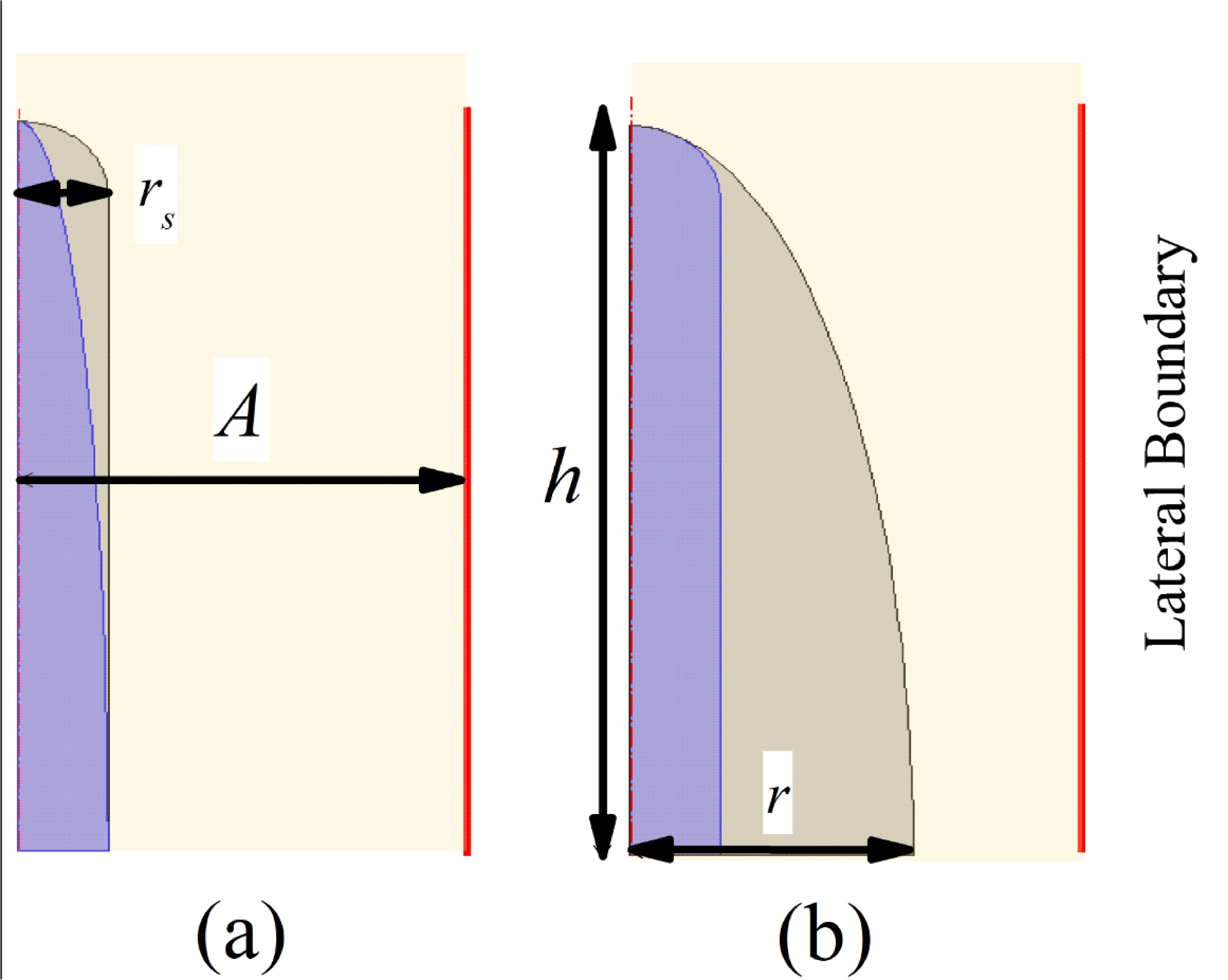}
\caption{Comparison between and HCP and an ellipsoidal emitter. In (a), with same aspect ratio, where the HCP experiences greater electrostatic influence from the lateral boundary. In (b), with same curvature at the apex, where the ellipsoidal emitter experiences greater influence from the boundary. In this case, the $\epsilon$ calculated for the ellipsoid is certainly going to be smaller for the HCP.} \label{hemihcpl}
\end{figure}

The HCP geometry is a classical representation of CNT emitters, so it is an important case to be analyzed here. For unitary aspect ratio, hemi-ellipsoids becomes hemispheres, so Eqs. (\ref{Ahmin}) and (\ref{Bhmin}) gives the MDS either for ellipsoid or HCP as an upper limit for the MDS. However, the improved MDS as a function of the aspect ratio does not scale for the HCP as it does for ellipsoids [Eqs. (\ref{Ahmin2}) and (\ref{Bhmin2})]. Note, in Fig. \ref{hemihcpl}(a) an HCP is superposed to an ellipsoid with same aspect ratio. The surface of the HCP is closer to the lateral boundary, so the boundary's influence is greater over the HCP then it is over the ellipsoid, implying that Eqs. (\ref{Ahmin2}) and (\ref{Bhmin2}) does not grant $\epsilon$ to be smaller than $\epsilon_{tol}$. Nevertheless, we can compare an HCP and an ellipsoid with same radius of curvature at the apex, as shown in Fig. \ref{hemihcpl}(b). In this case, the $\epsilon$ for the HCP is granted to be smaller than $\epsilon_{tol}$. Hence, we can safely use Eqs. (\ref{Ahmin2}) and (\ref{Bhmin2}) to evaluate the MDS for HCPs if we use the aspect ratio of the circumscribed ellipsoid, instead of the actual aspect ratio of the HCP. To do so, we must find the relation between the aspect ratio of the circumscribed ellipsoid as a function of the aspect ratio of the HCP.

Let $1/r_{s}$ be the curvature of the hemispherical cap that equals the curvature of the ellipsoid at the apex $1/r_e$ ($1/r_e=1/r_s$). The curvature ($1/r_e$) is simply the absolute value of the second derivative of the ellipsoid function $y=h\left(1 - x^{2}/r^{2} \right)$. Therefore:

\begin{equation}
 \frac{1}{r_e} = \left| h \frac{d^{2}}{dx^{2}} \left[ \sqrt{1 - \frac{x^2}{r^2} }\right]\right|_{x=0},
 \label{Curvature}
\end{equation}
which results in the following relation for the aspect ratios:

\begin{equation}
 \frac{h}{r_e} = \sqrt{\frac{h}{r_{s}}}.
 \label{Curvature2}
\end{equation}
Finally, we can improve the MDS for HCP as a function of the aspect ratio by replacing $h/r$ in Eqs. (\ref{Ahmin2}) and (\ref{Bhmin2}) for $\sqrt{h/r}$. As a last note, the procedure we did in section \ref{STFE} cannot be used to determine the MDS for HCPs, because, at present, there is not an analytical solution for the electrostatic potential distribution for single tip HCP emitters. However, the results presented here for this model are expected to significatively advances to obtain numerical values of $\gamma_{a}$ with unprecedented accuracy.

\subsection{Full 3D models of clusters}

To determine the MDS for cluster is somewhat more complicated. Clusters of emitters do not present rotational symmetry and cannot be simulated in a 2D axisymmetric model. Even then, Eqs. (\ref{Ahmin2}) and (\ref{Bhmin2}) still grants an error slightly smaller than the tolerance, because the boundaries in a full 3D model cause less electrostatic influence over the emitters, as explained in Ref.\cite{Fuzinato}. However, the distance $A$ from the center of the system to the lateral boundary is not a parameter of merit in clusters. Instead, we are interested in the minimal distance from the outmost emitter to the lateral boundary. To avoid confusion, let $C_{min}$ be the distance from the outer emitter to the boundary.

\begin{figure}[h!]
\includegraphics [width=8.0cm,height=5.8cm] {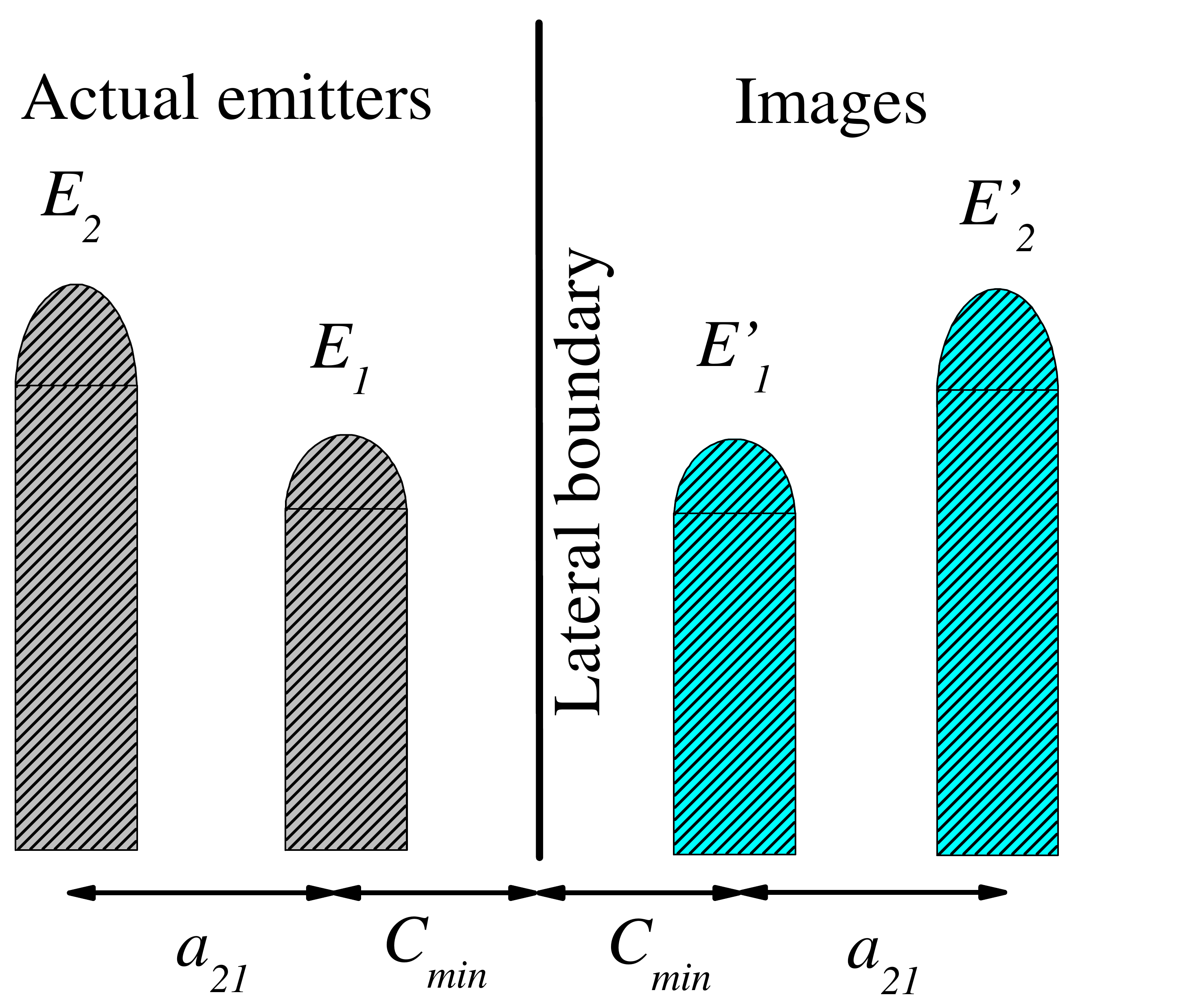}
\caption{Illustration of the images from a pair of emitters in a cluster. The $C_{min}$ that yields $\epsilon_{tol}$ is obtained from the electrostatic influence of the images on the outer emitter.} \label{hemihcplill}
\end{figure}

For clusters, the $\epsilon_{tol}$ at a particular emitter must be equal (in the condition for the MDS) to the summation of the errors caused by all emitters:

\begin{equation}
 \epsilon_{tol} = \sum\limits_{i=1}^{N} \epsilon_{i}.
 \label{errorcluster}
\end{equation}
where $N$ is number of emitter in the cluster. To evaluate $\epsilon_{i}$, we must be careful to consider the maximum influence that the boundary causes on the outmost emitter in the cluster. Consider only two emitter as shown in Fig. \ref{hemihcplill}. The image of emitter $E_1$ (indicated as $E'_{1}$) with respect to the lateral boundary generates an error in its own FEF, which is simply the inverse function of Eq. (\ref{Ahmin2}), given by:

\begin{equation}
 \epsilon_{1} \approx 174.7 \times \frac{0.2 + 0.8\exp\left[-0.345\left(h_{1}/r_{1} - 1 \right) \right]}{\left(C_{min}/h_{1}\right)^{3}}.
 \label{Ahmin2we}
\end{equation}
Parameters $h$ and $r$ does not need to be the same for all emitters. Therefore, these parameters gain a sub-index ($i=1,2,3...$), corresponding to the $i$-th emitter. The image of $E_{2}$ (indicated as $E'_{2}$) generates an error on $E_1$ according to the distance from $E_2$ to the lateral boundary plus the distance ($d_{21}$) from $E_2$ to $E_1$. Now it is important to note that Eqs. (\ref{Ahmin2}), (\ref{Bhmin2}) and (\ref{Ahmin2we}) assume implicitly that the image of the emitter is at a distance twice farther than the boundary. Therefore, to compute the error due to the influence of $E'_{2}$ on $E_1$, we must use half the distance between these (which is $C_{min}+a_{21}/2$), independently of the boundary's position. Hence, we have:

\begin{equation}
 \epsilon_{2} \approx 174.7 \times \frac{0.2 + 0.8\exp\left[-0.345\left(h_{2}/r_{2} - 1 \right) \right]}{\left[(C_{min}+a_{21}/2)/h_{2}\right]^{3}}.
 \label{Ahmin2we11}
\end{equation}
where $a_{21}$ is known. The influence of construction in Eq. (\ref{Ahmin2we11}) makes $\epsilon_2$ always larger than the error expected for $E_2$ isolated, which grants the total error to be lower than the tolerance. Replacing Eqs. (\ref{Ahmin2we}) and (\ref{Ahmin2we11}) in (\ref{errorcluster}), we get:

\begin{equation}
 \epsilon_{tol} \approx 174.7 \times\sum\limits_{i=1}^{2}\left\{ \frac{0.2 + 0.8\exp\left[-0.345\left(h_{i}/r_{i} - 1 \right) \right]}{\left[(C_{min}+a_{i1}/2)/h_{i}\right]^{3}}\right\},
 \label{Ahmin2wetot}
\end{equation}
where $a_{11}=0$. The $C_{min}$ in Eq. (\ref{Ahmin2wetot}) cannot be isolated, but can be solved straightforwardly using numerical methods (like Newton's method, for example). Finally, in a cluster with $N$ emitters, we just have to replace the upper limit of the summation in Eq. (\ref{Ahmin2wetot}) to $N$. Similarly, we can obtain the minimum distance to the top boundary as:

\begin{equation}
 \epsilon_{tol} \approx 99.9 \times\sum\limits_{i=1}^{N} \left\{\frac{0.2 + 0.8\exp\left[-0.345\left(h_{i}/r_{i} - 1 \right) \right]}{\left[(D_{min}+ h_{tall} + \Delta h_{i} )/h_{i}\right]^{3}}\right\},
 \label{Ahmin2wetotmin}
\end{equation}
where $D_{min}$ is the distance from the top of the tallest emitter to the top boundary, $h_{tall}$ is the height of the tallest emitter and $\Delta h_{i}$ is the difference between the tallest emitter and the $i$-th emitter. In Eq. (\ref{Ahmin2wetotmin}) we do not consider the relative positions amongst the emitter in the cluster, only the heights relative to the top boundary.

Remember, if the emitter in the clusters are HCPs, the aspect ration $hi/ri$, must be replaced by $\sqrt{hi/ri}$ as discussed.

\subsection{Table for reference and verification of the MSD method}

Table \ref{tabII} shows the $\left[(A/h)_{min},(B/h)_{min}\right]$ calculated from Eqs. (\ref{Ahmin2}) and (\ref{Bhmin2}) for a given tolerated error $\epsilon_{tol}$. Then, we verify the agreement between the tolerated error and the actual numerically evaluated error, $\epsilon_{num}$. As expected, the $\epsilon_{num}$ is always below $\epsilon_{tol}$, albeit, not by too much, as we do not want to overestimate the MDS.

\begin{table*}
   \centering
 \renewcommand{\arraystretch}{1.5}

 %\begin{tabular}{|c|}
 % \hline
 %$H=0.1$ \\ \hline

 % \end{tabular}
 \caption{MDS for hemi-ellipsoid single tip field emitters. Given the tolerated error in the apex FEF, the MDS is calculated from Eqs. (\ref{Ahmin2}) and (\ref{Bhmin2}), then the actual error $\epsilon_{num}(\%)$ is calculated from numerical simulation for comparison.}
 \begin{tabular}{|c|c|c|c|}

  \hline
 $\epsilon_{tol}(\%)$ & $(h/r=1) \left[(A/h)_{min},(B/h)_{min}\right] - \epsilon_{num}$ & $(h/r=5) \left[(A/h)_{min},(B/h)_{min}\right] - \epsilon_{num}$ & $(h/r>20) \left[(A/h)_{min},(B/h)_{min}\right] - \epsilon_{num}$ \\

 \hline
 \hline

$10$ & $\left[2.59,2.15\right] - 9.40$ & $\left[1.91,1.59\right] - 6.41$ & $\left[1.52,1.26\right] - 8.03$    \\ \hline
$1$ & $\left[5.59,4.64\right] - 1.00$ & $\left[4.12,3.42\right] - 0.60$ & $\left[3.28,2.72\right] - 0.62$   \\ \hline
$0.1$  & $\left[12.04,10.00\right] - 0.10$ & $\left[8.88,7.37\right] - 0.0599$ & $\left[7.06,5.86\right] - 0.062$   \\ \hline\
$0.01$  & $\left[25.95,21.54\right] - 0.01$ & $\left[19.14,15.89\right] - 0.00596$  & $\left[15.20,12.62\right] - 0.0063$  \\ \hline
$0.001$ & $\left[55.9,46.4\right] - 9.4\times10^{-4}$ & $\left[41.23,34.22\right] - 5.55\times10^{-4}$  & $\left[32.75,27.19\right] - 6.8\times10^{-4}$    \\ \hline
$0.0001$ & $\left[120.4,99.96\right] - 3.7 \times 10^{-5}$ & $\left[88.83,73.73\right] - 1.4 \times 10^{-5}$  & $\left[70.56,58.57\right] - 1.2 \times 10^{-5}$   \\ \hline

\end{tabular}
%\vspace{0.5cm}

\label{tabII}
\end{table*}

\section{Conclusions}
\label{Conc}

We provided the MDS of the simulation domain $(A/h)_{min}$ and $(B/h)_{min}$ necessary to simulate the apex-FEF with a systematic error smaller than a given $\epsilon$. We summarize the results as follows:

(i) For a hemi-ellipsoid single tip field emitter (STFE), with a given aspect ratio ($h/r$), the MDS is

\begin{equation}
 (A/h)^{STFE}_{min} \approx 5.59 \times \sqrt[3]{\frac{0.2 + 0.8\exp\left[-0.345\left(h/r - 1 \right) \right]}{\epsilon_{tol}}},
 \label{Ahmin2C}
\end{equation}
and
\begin{equation}
 (B/h)^{STFE}_{min} \approx 4.64 \times \sqrt[3]{\frac{0.2 + 0.8\exp\left[-0.345\left(h/r - 1 \right) \right]}{\epsilon_{tol}}}.
 \label{Bhmin2C}
\end{equation}

(ii) For a planar hemi-ellipsoid capacitor system, $B_{Anode}/h$ is fixed and

\begin{equation}
 (A/h)^{capacitor}_{min} \approx 5.59 \times \sqrt[3]{\frac{0.2 + 0.8\exp\left[-0.345\left(h/r - 1 \right) \right]}{\epsilon_{tol}}}.
 \label{Ahmin2CC}
\end{equation}

(iii) For an infinite array formed by hemi-ellipsoid emitters, $A_{array}/h$ is fixed and

\begin{equation}
 (B/h)^{array}_{min} \approx 4.64 \times \sqrt[3]{\frac{0.2 + 0.8\exp\left[-0.345\left(h/r - 1 \right) \right]}{\epsilon_{tol}}}.
 \label{Bhmin2CC}
\end{equation}

(iv) For HCP emitters, the MDS is the same as in Eqs. (\ref{Ahmin2C}), (\ref{Bhmin2C}), (\ref{Ahmin2CC}) and (\ref{Bhmin2CC}), except that the aspect ratio $h/r$ must be replaced to $\sqrt{h/r}$.

To simulate isolated emitters, arrays or clusters of emitters, our results are only valid when using Neumann BC at top and right hand side boundaries, which the authors strongly recommend. The Neumann BC generates systematic errors much smaller than the Dirichlet BC for a given domain size; typically by a factor of $10$. In other words, the MDS using a specified voltage at the contours of the simulation domain should be bigger to yield the same precision.
We expect our results be a major assistance for the field emission community to obtain accurate FEF in systems where the analytical electrostatic solution is still unknown.

\section{Acknowledgements}
TAdA acknowledges Royal Society under Newton Mobility Grant, Ref: NI160031.  The authors are thankful to the Brazilian Council of Science and Technology (CNPq) and to Richard Forbes for his suggestions and discussions.
%\bibliography{SCbib3N}



\end{document}